# Cavity piezooptomechanics: piezoelectrically excited, optically transduced optomechanical resonators

Chi Xiong, Linran Fan, Xiankai Sun, and Hong X. Tang[a]

*Department of Electrical Engineering, Yale University, 15 Prospect St., New Haven, Connecticut 06511, USA*

**Abstract:** We present a monolithic integrated aluminum nitride (AlN) optomechanical resonator in which the mechanical motion is actuated by piezoelectric force and the displacement is transduced by a high-$Q$ optical cavity. The AlN optomechanical resonator is excited from a radio-frequency electrode via a small air gap to eliminate resonator-to-electrode loss. We observe the electrically excited mechanical motion at 47.3 MHz, 1.04 GHz, and 3.12 GHz, corresponding to the $1^{st}$, $2^{nd}$, and $4^{th}$ radial-contour mode of the wheel resonator respectively. An equivalent circuit model is developed to describe the observed Fano-like resonance spectrum.

---

[a]Electronic mail: hong.tang@yale.edu.

Cavity optomechanical resonators have exhibited ultrasensitive displacement sensitivity close to the quantum limit[1] and hold promise for technological advances in metrology and quantum information processing[2]. So far, optomechanical actuation has been mostly achieved via radiation pressure effects[3-6], which are inherently weak[7-9] and often give rise to deleterious heating effects.[10] Recently cavity optoelectromechanical resonators,[11-13] which combine stronger electrostatic actuation and ultrasensitive displacement readout by an optical cavity, have been demonstrated. Electrical actuation of the optomechanical resonators could enable feedback cooling or amplification of mechanical resonators for realizing low phase noise oscillators.

In micro- and nanoelectromechanical systems (MEMS/NEMS), piezoelectric force[14, 15] can provide strong mechanical excitation besides electrostatic forces. Compared with micromechanical resonators using electrostatic effects, piezoelectric transduction is particularly attractive for ultrahigh-frequency applications, such as thin-film bulk acoustic resonators (FBAR)[16] and contour-mode resonators[14]. In addition to piezoelectric force actuation, conventional piezoelectric resonators usually rely on metal electrodes in contact with the device to enable displacement readout by the piezoelectric effect, which transduces the displacement to a sensing voltage. Compared to electrical readout, optical sensing offers superior sensitivity and simplifies the fabrication of the device. Recently we have demonstrated an AlN cavity optomechanical resonator in a shape reminiscent of a wheel with displacement sensitivity as high as $6.2 \times 10^{-18}$ m/√Hz[17]. The Brownian motions of the wheel resonator were transduced optically. In this Letter, we show that such AlN optomechanical resonators can be excited more efficiently by piezoelectric force and we observe driven response of the mechanical resonances at 47.3 MHz, 1.04 GHz, and 3.12 GHz, corresponding to the 1st, 2nd and 4th radial-contour mode of the wheel respectively. The measured mechanical $Q$ factors are 1370, 2470, for the 47.3 MHz and 1.04 GHz mode in ambient air respectively. Our work points to a direction for building a class of piezooptomechanical systems with an additional electrical degree of freedom.

The AlN wheel resonators under study are fabricated in 330-nm-thick *c*-axis-oriented AlN thin films sputtered on oxidized silicon wafers. The fabrication details were described elsewhere.[18] Fig. 1(a) shows the measurement setup. Light from a tunable diode laser (TDL) is coupled into and out of the device using a pair of grating couplers. A fiber polarization controller (FPC) is used before the input grating coupler to select the TE-like polarization. The transmitted light is detected by a low-noise InGaAs photoreceiver (PR). The piezoelectric transduction usually requires the devices to be fabricated with metal electrodes in physical contact. Here in order to avoid the optical loss induced by the metal, we separate the wheel resonator from the electrode via a small air gap (about 50 µm), as shown in the setup diagram. The electrode is made from a sharpened stainless steel probe with a tip diameter of about 10 µm. This contactless design simultaneously serves to eliminate the mechanical loss from the resonator–electrode contact while maintaining high electromechanical coupling.[19]

Fig. 1(b) shows the optical transmission spectrum of a typical wheel resonator which shows a loaded optical $Q$ factor of 125, 000. The inset shows a scanning electron micrograph of the AlN wheel (inner radius $R_i = 32.6$ µm, outer radius $R_o = 37.6$ µm) and the zoom-in of an optical resonance near 1544.85 nm. As shown in the schematic of Fig. 1(a), we position the electrode in close proximity vertically above the center of the wheel, while the silicon substrate is connected to ground. Optical transduction of the mechanical modes is performed by tuning the input laser to the wavelength corresponding to the maximum slope of an optical resonance. We keep the laser input power low enough so that the optomechanical back-action is negligible. The transmitted light is amplified by an erbium doped optical amplifier (EDFA) and then detected by a high-speed InGaAs photoreceiver with 4 GHz bandwidth.

The dynamic response of the AlN wheel resonator under an AC drive is characterized by measuring $S_{21}$ transmission spectrum from the network analyzer. Fig. 2 shows a wideband spectrum of the measured magnitude and phase of $S_{21}$ with zero DC bias voltage. We observe three dominant peaks near 47.3 MHz, 1.04 GHz, and 3.12 GHz, corresponding to the 1st, 2nd and 4th radial-contour mode respectively. With predominant in-plane motions, these modes are efficiently excited through the the off-diagonal component ($d_{13}$) of AlN's piezoelectric coefficient tensor under the present electrode configuration, which provides a large out-of-plane ($E_z$) field component. We do not observe the 3rd radial-contour mode expected to be near 2.1 GHz, which can be attributed to the symmetry of the displacement profile and the vanishing overlap integral with the vertical excitation field. We tune the DC bias voltage from 0 up to 20 V and do not observe any change in the signal amplitude, consistent with the properties of piezoelectric actuation. The frequencies of the respective modes agree with those obtained from theoretical and numerical calculations.[17]

The mechanical resonances exhibit Fano-like asymmetric line-shape as shown in Fig. 2 (a). Generally, Fano resonance is a result of interference between a resonant process and a slowly varying background such as nonresonant signal crosstalk. In our case, since the direct electrical crosstalk is negligible due to the optical isolation, the electrical–optical–electrical crosstalk originates from the Pockels eeffect[17] in the AlN wheel. The cavity resonance shift ($\Delta f$) can be written as a sum of the effects from Pockels modulation ($\Delta f_{eo}$) and optomechanical modulation ($\Delta f_{om}$):

$$\Delta f = \Delta f_{eo} + \Delta f_{om} \quad (1)$$

To derive the optomechancial modulation, we model the air gap–AlN–air gap stack by three capacitors in series, as shown in the equivalent circuit in Fig. 1(b). Adopting the Butterworth-van Dyke (BvD) equivalent circuit model, the mechanical resonator is represented as a motional impedance of $Z_m$, which consists of a series of motional resistance $R_m$, motional capacitance $C_m$, and motional inductance $L_m$. The displacement of the mechanical resonator is proportional to the "charge" stored in the motional capacitor $C_m$. The displacement $x$ can be calculated as

$$x = \frac{\eta V}{\left(1 + C_p/C'\right)\left(i\omega R_m - \omega^2 L_m + 1/C_m + 1/(C' + C_p)\right)} \tag{2}$$

where $C'$ is the equivalent series capacitance of $C_1$ and $C_2$: $C' = \dfrac{C_1 C_2}{C_1 + C_2}$, $\eta$ is a constant and $\omega$ is the actuation frequency. Considering a Pockels modulation coefficient of $b$ (GHz/V) and an optomechanical coupling coefficient of $g$ (GHz/nm), we can derive the total cavity resonance modulation ($\Delta f$) as

$$\begin{aligned}
\Delta f &= \Delta f_{eo} + \Delta f_{om} \\
&= bV + gx \\
&= bV + \frac{g\eta V}{\left(1 + C_p/C'\right)\left(i\omega R_m - \omega^2 L_m + 1/C_m + 1/(C' + C_p)\right)}
\end{aligned} \tag{3}$$

The intensity modulation $I_{AC}$ can be derived as

$$I_{AC}(\omega) = |\Delta f| = V \left\{ \frac{\left[g\eta + b/C' + b(1 + C_p/C')(\omega^2 L_m - 1/C_m)\right]^2 + \omega^2 (1 + C_p/C')^2 R_m^2 b^2}{\left[1/C' + (1 + C_p/C')(\omega^2 L_m - 1/C_m)\right]^2 + \omega^2 (1 + C_p/C')^2 R_m^2} \right\}^{1/2} \tag{4}$$

On resonance $\omega = \omega_r$, we can simplify Eq. 4 and derive the intensity modulation ($I_{AC}$) as

$$I_{AC}(\omega = \omega_r) \approx V \cdot \frac{b(1+C_p/C')\left(\omega_r^2 + \frac{g\eta C'}{b(C_p+C')L_m}\right)}{\omega_r^2(1+C_p/C')} \cdot \frac{\left(4(\omega-\omega_r+q)^2 + \frac{R_m^2}{2L_m^2}\right)}{\left(4(\omega-\omega_r)^2 + \frac{R_m^2}{2L_m^2}\right)}$$

$$= C \cdot \frac{\left(4(\omega-\omega_r+q)^2 + \frac{R_m^2}{2L_m^2}\right)}{\left(4(\omega-\omega_r)^2 + \frac{R_m^2}{2L_m^2}\right)}$$

(5)

where $C$ is a constant, $\omega_r$ is the angular mechanical resonant frequency which is defined as $\omega_r = \sqrt{1/L_m C_m + 1/L_m(C'+C_p)}$, $q$ is the angular frequency difference between resonance and anti-resonance of the Fano-like line-shape, which is defined as $q = \sqrt{\omega_r^2 - g\eta/b(1+C_p/C')} - \omega_r$. We can see from Eq. 5 that the intensity modulation ($I_{AC}$) has a Fano-like line-shape. The fitting of the measurement data with the above Fano-like function reveals an excellent agreement, as indicated in Fig. 2(a), where the Fano-like fitting for the 1.04 GHz resonance is marked as the blue line. The fitting parameters for the 1.04 GHz mode are $f_r = \omega_r/2\pi = 1.04$ GHz, $Q_m = 2470$, and $q/2\pi = 130$ MHz. For the 47.3 MHz mode, the fitting parameters are $f_r = \omega_r/2\pi = 47.3$ MHz, $Q_m = 1370$, and $q/2\pi = 25.4$ MHz. The $Q$ values of the 47.3 MHz and 1.04 GHz modes are consistent with the results from our previous thermomechanical noise measurement[17], confirming that the contactless driving scheme does not degrade the mechanical quality of the resonator. We do not perform the fitting on the 3.12 GHz mode because this mode sees two closely spaced peaks when zoomed in most likely due to mode hybridization.

From the above derivations, we can extract the embedded Lorentzian mechanical resonance line-shape and show in Fig. 3(a)–(c) the magnitude and phase of the processed spectra of the 47.3 MHz, 1.04 GHz, and 3.12 GHz mode respectively, along with the corresponding numerically calculated cross-sectional mechanical displacement profiles. The displacement is calculated by comparing the measured voltage noise spectral density and the displacement noise spectral density and then deducing the transduction gain (V/m) of the device. A large enhancement of the mechanical displacement is observed as the RF drive signal approaches the mechanical resonant frequency. For the 2nd radial-contour mode at 1.04 GHz, we also measure the spectra under different driving power from −30 dBm to 0 dBm. The on-resonance displacement as a function of the *rms* voltage applied on the electrode is plotted in Fig. 3(b) inset, showing a linear relationship which indicates that the resonator still operates in the linear regime. For the 47.3 MHz and 3.12 GHz modes, the excitation power is fixed at 0 dBm.

The magnitude of the piezoelectric driving force ($F_{pz}$) can be calculated by $F_{pz} = k\,x(\omega_r)/Q_m = \omega^2 m_{eff} x(\omega_r)/Q_m$, where $k$ is the spring constant of a mechanical mode, $x(\omega_r)$ is the mechanical displacement on resonance and $Q_m$ is the mechanical quality factor. For the 1.04 GHz mode, the effective mass is 0.42 ng, the on-resonance displacement is $3 \times 10^{-11}$ m for RF power of 0 dBm (1 mW), the piezoelectric force at this RF power can then be calculated as 0.22 µN. We can further compare the magnitude of the piezoelectric force to that of the optical force and the electrostatic force in our system. The optical force can be calculated as $F_o=|a|^2 g_{om}/\omega$, where $\omega$ is the optical cavity frequency ($2\pi \times 194$ THz), $g_{om}$ is the optomechanical coupling coefficient ($g_{om} \sim \omega/R = 2\pi \times 6.0$ GHz/nm). The intra-cavity field intensity $|a|^2$ can be calculated as

$$|a|^2 = \frac{2\kappa_c P}{\kappa^2 + \Delta^2}$$

, where $\kappa_c$ is the cavity dissipation rate ($2\pi \times 0.5$ GHz), $\kappa$ is the cavity coupling rate ($2\pi \times 1$ GHz), $\Delta$ is the cavity detuning ($2\pi \times 1$ GHz), P is the input optical power (1 mW). After plugging the values of each of the quantities, we obtain the optical force as $F_o$=2.5 nN. The electrostatic dipole force in our system can be estimated by $F_{el} = \frac{1}{2}\frac{dC}{dg}V_{DC}^2$ when applying a DC voltage $V_{DC}$ over a capacitor $C$ and a gap of $g$. For an electrode to device gap of 50 µm we estimate C~0.2 fF and $F_{el}$~1.9 pN/$V_{DC}^2$. It is evident that the piezoelectric force is the dominant force in our system.

In conclusion, we have demonstrated a monolithic piezoelectrically actuated aluminum nitride optomechanical resonator operating at frequencies up to 3.12 GHz. By taking advantage of the piezoelectric driving force as large as 0.22 μN and high displacement sensitivity offered by a high-$Q$ optical cavity, the radial-contour modes of the wheel resonators are efficiently excited and optically read out. Using on-chip electrodes in future experiments could further enhance the excitation efficiency and provide access to study the nonlinear dynamics of optomechanical resonator under strong piezoelectric drive. These AlN piezoelectric optomechanical resonators can be employed to build low-switching-voltage optical modulators[11] and low-phase-noise opto-acoustic oscillators.[20] Our piezo-opto-mechanical system can also find applications for optomechanical cooling and amplification by using electrical feedback.

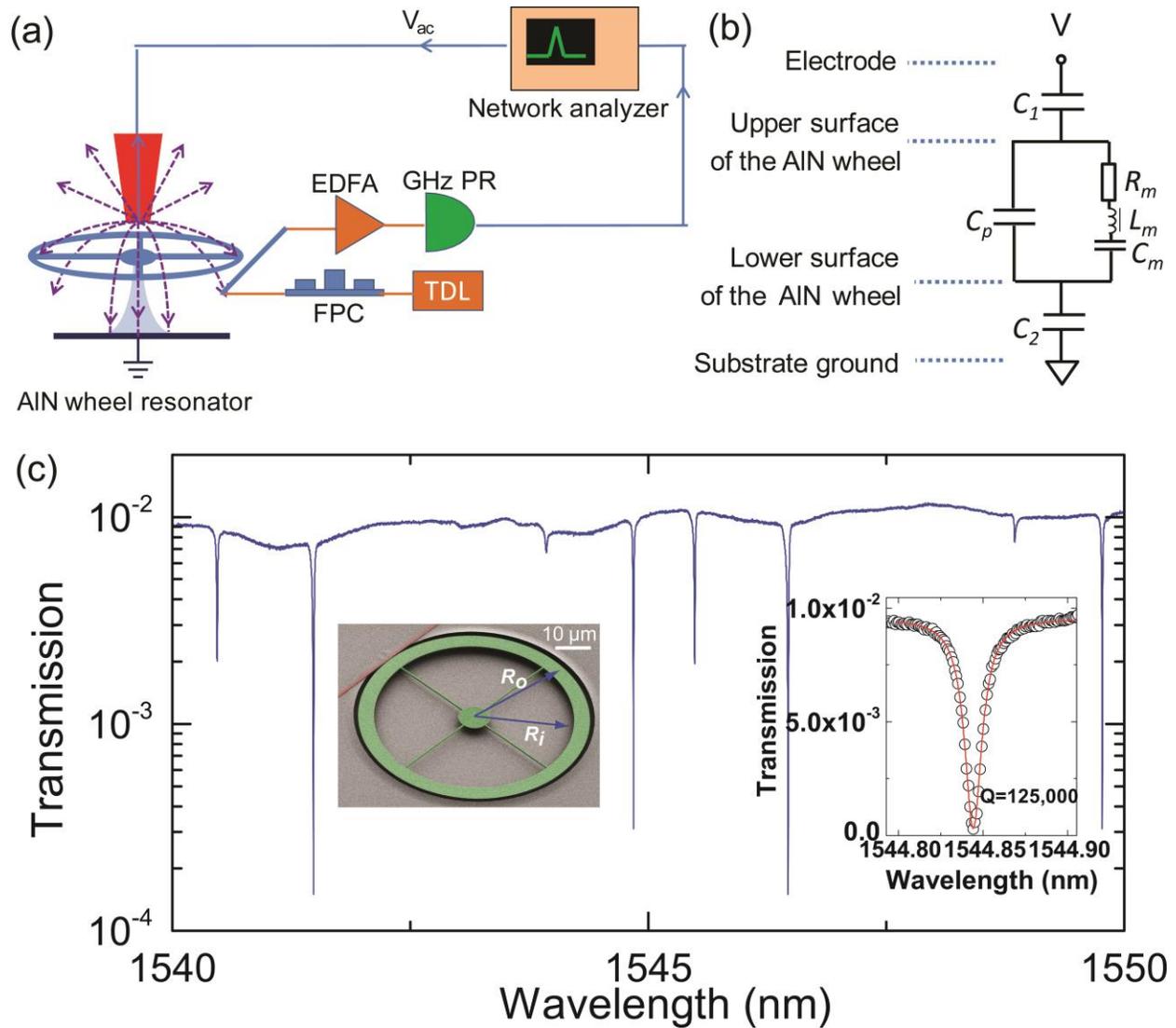

FIG. 1. (a) Setup for measuring the optical response of the piezoelectrically actuated AlN optomechanical resonator. EDFA: erbium-doped fiber amplifier. PR: photoreceiver. FPC: fiber polarization controller. TDL: tunable diode laser. The RF voltage from a network analyzer is applied to the electrode placed directly above the center of the wheel resonator. The silicon substrate is grounded and the electric field lines are shown in dotted purple lines. (b) The equivalent circuit of the air gap–AlN–air gap resonator stack modeled as three capacitors in series. $R_m$, $L_m$, $C_m$ are the motional resistance, inductor, and capacitance of the mechanical resonator respectively. $C_1$ is the physical capacitance between the electrode and the wheel's

upper surface, $C_2$ is the physical capacitance between the wheel's lower surface and the ground and $C_p$ is the dielectric capacitance of the AlN device layer. (c) Typical optical transmission of an AlN wheel optomechanical resonator with the inner radius $R_i$ = 32.6 μm and outer radius $R_o$ = 37.6 μm. The left inset is a scanning electron micrograph of the AlN wheel resonator and the right inset is the zoom-in of a resonance near 1544.85 nm with a loaded optical $Q$ factor of 125,000.

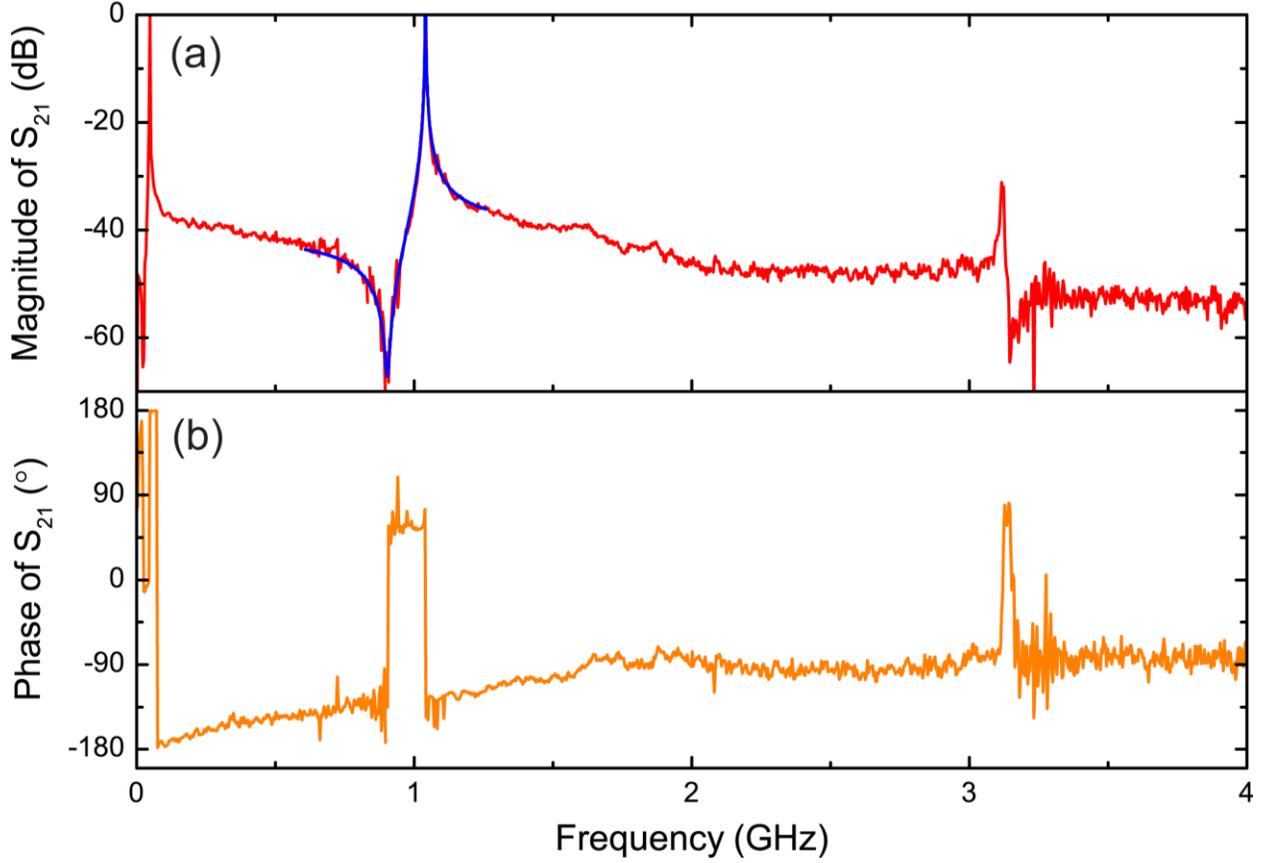

FIG. 2. Spectrum of magnitude (a) and phase (b) of the $S_{21}$ measurement from 0 to 4 GHz showing three distinct modes around 47.3 MHz, 1.04 GHz, and 3.12 GHz. The asymmetric line-shape of the magnitude spectrum around the resonances can be modeled as a Fano-like resonance as a result of interference between the mechanical resonances and Pockels modulation. The blue line around 1.04 GHz is a Fano-like line-shape fitting to Eq. 5 with the following fitted parameters: $\omega_r/2\pi$ = 1.04 GHz, $Q_m$ = 2470, and $q/2\pi$ = 130 MHz.

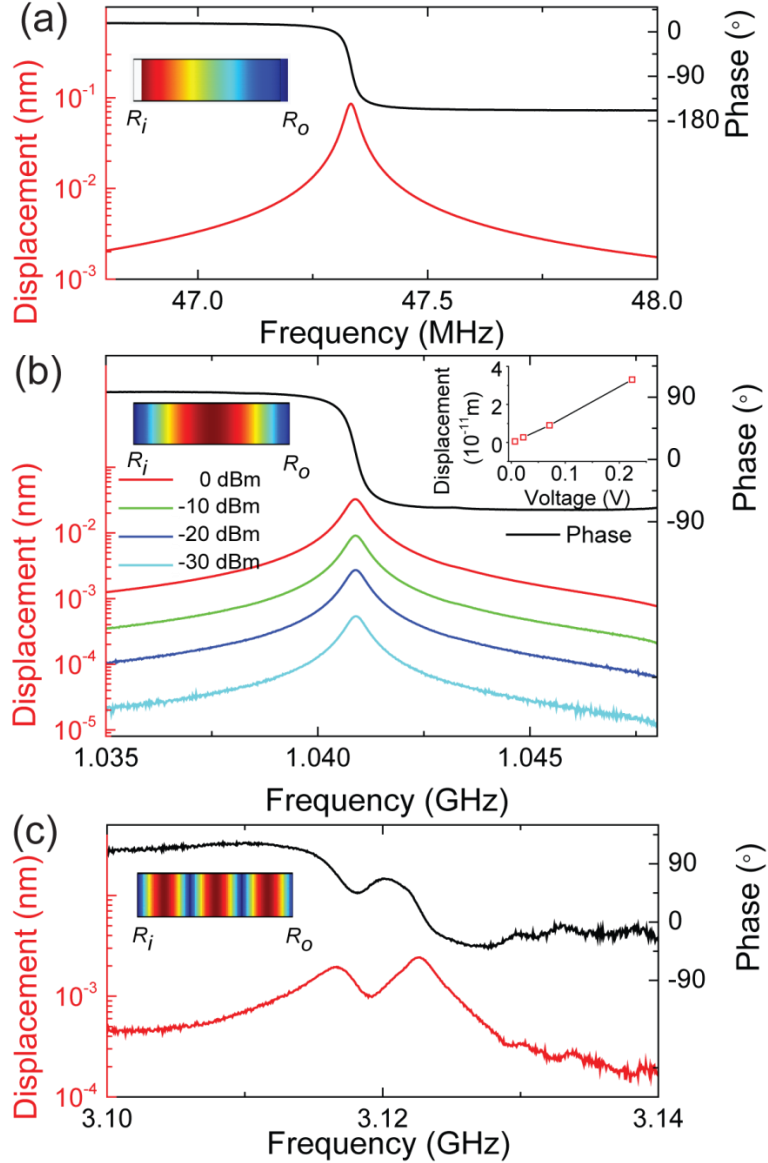

FIG. 3. (a) Zoomed-in displacement amplitude and phase spectrum of the mechanical resonance around 47.3 MHz. (b) Zoomed-in spectrum of the mechanical resonance around 1.04 GHz under various driving power levels from −30 dBm to 0 dBm. The plot in the right inset shows a linear dependence of the displacement on the driving voltage. (c) Zoomed-in spectrum of the mechanical resonance around 3.12 GHz. The numerically calculated cross-sectional displacement mode profiles in the radial direction are shown in the left inset respectively for each mechanical mode.